\documentclass[twocolumn]{aastex63}

\usepackage{amsmath}
\newcommand{\hi}{\ifmmode{\rm HI}\else{H\/{\sc i~}}\fi} 

\graphicspath{{./}{figures/}}
\shorttitle{Burton's Curse}
\shortauthors{Peek et al.}

\begin{document}

\title{Burton's Curse: The Impact of Bulk Flows on the Galactic Longitude--Velocity Diagram and the Illusion of a Continuous Perseus Arm}

\correspondingauthor{J. E. G. Peek}
\email{jegpeek@stsci.edu}
\author[0000-0003-4797-7030]{J. E. G. Peek}
\affiliation{Space Telescope Science Institute, 3700 San Martin Drive, Baltimore, MD 21218, USA}
\affiliation{Department of Physics \& Astronomy, Johns Hopkins University, Baltimore, MD 21218, USA}
\author[0000-0003-0789-9939]{Kirill Tchernyshyov}
\affiliation{Department of Astronomy, University of Washington, Seattle, WA, USA}
\author[0000-0002-7351-6062]{Marc-Antoine Miville-Deschenes}
\affiliation{AIM, CEA, CNRS, Université Paris-Saclay, Université Paris Diderot, Sorbonne Paris Cité, F-91191 Gif-sur-Yvette, France}

\begin{abstract} In this work we demonstrate that the Perseus Arm is not a continuous structure of molecular gas in the second quadrant. We first show that the observed, distanced-resolved velocity structure of the Galaxy in the outer disk is capable of creating illusory spiral arms, as was first proposed by \cite{1971A&A....10...76B}. Second, we measure the distances to a collection of CO clouds at velocities consistent with the Perseus arm with $135^\circ < l < 160^\circ$. We find these distances using 3D dust maps from \cite{Green:go}. We determine that these molecular cloud do not preferentially lie at the distance of a purported Perseus arm, but rather extend over 3 kpc in distance, with some evidence for a closer, high pitch angle structure between 1 and 1.5 kpc away. Finally, we demonstrate that velocity perturbations of the amplitude found near the Perseus arm can wreak havoc on our interpretation of the longitude--velocity diagram for more than half of the Milky Way disk.

\end{abstract}


\section{Introduction} \label{sec:intro}

Much of the star formation in disk galaxies happens in their spiral arms, and thus understanding spiral structure is key to understanding the process of star formation on large scales. On the other hand, we know star formation to be mediated by very small-scale stellar feedback and turbulent processes, which can be studied in most detail within the Milky Way. Thus, understanding the true nature of Milky Way spiral structure is critical to connecting the environment of star formation to its eventual result.

While our need to accurately map Milky Way spiral structure has long been recognized, our location within the Milky Way disk limits our ability to discern the distances to the molecular clouds and star forming regions that make up spiral arms. Emission tracers of the ionized, atomic, and molecular gas that make up the arms contain no intrinsic distance information, and stellar structures embedded within the arms are often obscured by intervening material. Even \emph{Gaia}'s state of the art, large scale precise parallactic distance information to individual stars is  largely limited to nearby ($\lesssim$ 2 kpc), relatively unobscured, and unconfused regions \cite{Collaboration:io}. Very Long Baseline Array (VLBI) parallax measurements toward high-mass star formation regions (HMSFRs) provides a much more limited set of stellar distances, but one that is unaffected by dust, and traces the star forming regions of interest \citep[][Reid14]{Reid:2014km}.

Historically, astronomers have estimated distances to gas seen in emission by taking advantage of the analytic relationship between distance and radial velocity that is created by the differential rotation of the Milky Way disk \citep{1954BAN....12..117V}. In particular, in the outer Milky Way there is a one-to-one, monotonic relationship between distance and velocity that, under the assumption of a known and unperturbed rotation curve, allows us to exactly map out the structure of the disk beyond the solar circle (\cite{Levine2006}, \cite{kalberla09}, \cite{2017ApJ...834...57M}). This method represents one of the only ways to connect individual regions of star formation (as traced by e.g. HMSFRs) into larger spiral structures across the Galactic disk. This linking is very important, as the pitch angle and continuity of spiral structure is a key discriminant between different theories of spiral arm formation and persistence (Reid14). 

\citet[B71]{1971A&A....10...76B} examined this velocity-based map-making method and pointed out a fundamental flaw. By modeling a \emph{completely smooth} ISM with the velocity field implied by the \citet{1969ApJ...155..721L} spiral structure density wave theory, B71 was able to reproduce many of the observed features in the Galactic \hi longitude--velocity diagram. The implication of this work is that realistic spiral structure \emph{velocity fields} generate perturbations to the longitude--velocity diagram that mimic in shape and amplitude those generated by spiral arm density fluctuations themselves. \cite{Burton:1974wo} followed up on this analysis, comparing distances to star clusters to the implied distances from velocities in \hi . While these original theories of spiral structure are by no means the only ones under consideration for the Milky Way, recent work has shown that the Milky Way does indeed host line-of-sight velocity fluctuations of more than 10 km/s stretching for many kpc, which may be capable of generating these kinds of distortions \citep[][TPZ18]{Tchernyshyov:2017hw, Tchernyshyov:2018bk}.

The Perseus arm in the outer Milky Way has long been used as a test bed for theories of spiral structure, as it is (a) a purported nearby spiral arm, (b) in the outer Galaxy, (c) easily visible in the north, and  (d) seemingly continuous across a large swath of the Galactic disk. Early work argued that the velocity and density field in this region was consistent with the two-arm spiral shock model of spiral structure \citep{Roberts:1972bp}.
Subsequent observations of the positions and motions of gas and stars in the Perseus arm have also been found to agree with the \citet{Roberts:1972bp} model \citep{Burton:1974wo,Humphreys:1976iz,Sparke:1978fe,Wouterloot:1985wq,Sakai:2019bt}, though often with caveats.
\citet{Humphreys:1976iz} find that the line-of-sight velocities of supergiant stars have the right signs, but twice the expected amplitude. 
\citet{Sparke:1978fe} note the Perseus arm gap discussed in this paper, but conclude that it could in fact be the result of a non-linear effect hypothesized to occur in the spiral shock model. 
\citet[][Sakai19]{Sakai:2019bt} find that the 3D velocities of HMSFRs in the Perseus arm are consistent with the \citet{Roberts:1972bp} model but that HMSFRs in the rest of the purported spiral arms are not consistent with spiral shock-like models. 
The development of other theories of spiral structure continued, in particular variations on swing amplification \citep{Toomre:1981uv,Sellwood:1984fh}.
Unlike \citet{Roberts:1972bp}, these models did not make the sort of explicit predictions for the Perseus arm and so were typically not compared with observations. 

TPZ18 showed that the velocity fields around Perseus star-forming regions are inconsistent with these kinds of long-lived persistent spiral models, and that material arm models, in which spiral arms are created and destroyed on less than a winding time, are more consistent with the observed velocity fields. These material arm models are incapable of making long, shallow pitch angle spiral arms, which stands in contradiction to the observed length and tight winding in the Perseus arm as identified in the longitude--velocity diagram. Here, shallow pitch angle refers to the small angle made by the arm and a constant radius circle \citep[see e.g.][]{SM21}. New observations of hot stars, Cepheids, and open clusters mined from {\em Gaia} EDR3 \citep{Collaboration:ev} also seem to show a more loosely wound Perseus arm \citep{Poggio:wd}. Longer, more tightly wound spiral arms, as argued for in Reid14, rely on sparse HMSFRs linked together by features in the molecular (CO) and atomic (H\/{\sc i}) longitude--velocity diagrams. While the distances and velocities of HMSFRs are not in doubt, can we really trust the links between them, or are they an illusion, as described in B71?

In this work, we put the Perseus arm to the test, examining whether a key part of this arm is indeed contiguous, or rather an illusion generated by Galactic velocity fields as described in B71. The part of Perseus we examine here does not have any known masers, but is important in connecting up star forming regions in the second and third quadrants, and thus establishing the shallow pitch angle of the purported arm. In Section \ref{sec:data} we describe the 3D dust maps, CO, H\/{\sc i}, and HMSFR maser data we use. In Section \ref{sec:meth_anal} we show how previously measured radial velocity fields provide for the possibility of spiral arm illusions in the Perseus arm, and describe our method for measuring the distance to individual molecular clouds in this region. In Section \ref{sec:disc} we discuss the implications of our measured distance distribution of molecular clouds in the region under consideration and we conclude in Section \ref{sec:conc}.

\section{Data} \label{sec:data}

\subsection{3D dust maps}
We use the 3D dust data products described in \citet[G19]{Green:go}. In brief, G19 uses the optical and near IR photometry of 799 million stars, along with their \emph{Gaia} data release 2 \cite{Collaboration:io} parallax measurements, to infer the combinations of stellar type, distance, and foreground reddening that are possible for each star. By combining these probability distribution functions for all stars near each line of sight, G19 produced a 3D dust map, with typical angular resolution 3.4$^\prime$, and distance bins of 0.125 distance moduli. We will refer to the total reddening profile along the line of sight from this map as ${\rm E\left(B-V\right)}\left(d\right)$ and to the radial differential reddening profile as $\Delta {\rm E\left(B-V\right)}\left(d\right)$. Adjacent bins are somewhat correlated, as a smooth spatial prior is applied in an iterative scheme to the map. Five realizations of the map are provided and we access them using the \emph{dustmaps} code \citep{dustmaps}. Additionally, each pixel in the map is labeled with its maximum and minimum reliable distance; lines of sight with few stars or opaque clouds cannot reliably return dust information beyond a certain point.

We note that G19 finds that the Reid14 HMSFRs are associated with regions of dense dust, but that there is no clear evidence for  the kind of shallow pitch angle spiral structure found in Reid14 in the G19 dust maps themselves. We believe this result is consistent with what we find in this work.

\subsection{Kinetic Tomography}

In order to show that the true velocity field of the Milky Way is capable of producing the significant velocity crowding effects as described in B17 we use the kinetic tomography map published in TPZ18. TPZ18 used the 1.53 $\mu$m diffuse interstellar band absorption feature in stars from APOGEE \citep{Majewski:ip, 2015ApJ...798...35Z} to measure the velocity field across the Milky Way disk in the first three quadrants within about 4 kpc. These maps confirmed the overall results of the previous maps \citep[][TP17]{Tchernyshyov:2017hw}, which showed that there are bulk flows of material in the Galaxy that can stretch many kpc at tens of km s$^{-1}$ beyond what is expected from Galactic rotation.  Additionally, we use the results of TP17 as an illustration of our results in Section \ref{sec:disc}.

\subsection{ISM gas tracer observations}

We use the compilation of 21-cm Hydrogen (\hi ) and microwave carbon monoxide (CO) observations described in TP17 to provide a template for contaminating reddening, as described in \S \ref{sec:bc_is_real}. Practically speaking, this means we used CO observations from \citet[][DHT01]{2001ApJ...547..792D} and \hi  observations from \cite{Kalberla05}.

\subsection{Cloud identification}

To select individual molecular clouds, we used the decomposition of the DHT01 map provided by \cite[MML17]{2017ApJ...834...57M}. MML17 used a hierarchical cluster identification method to collect together Gaussian velocity components of the DHT01 CO survey into a list of 8246 molecular clouds across the Galactic disk. The data contain both a list of the 8246 clouds and their bulk properties, but also a list of the individual Gaussian components associated with each cloud. This augmented and slightly modified MML17 data set is described in Appendix A and can be found at\dataset[10.7910/DVN/QR9CFW]{https://doi.org/10.7910/DVN/QR9CFW}.

\subsection{High Mass Star Formation Regions}

Finally, we also use the HMSFRs compiled in Reid14 and Sakai19. This represents the largest single collection of HMSFRs with parallactic distances ever assembled. These targets are very helpful for studying the global structure of star formation in the Milky Way because they are found in areas of active star formation and contain masers, whose parallaxes (and hence distances) and radial velocities can be measured accurately using VLBI. This combination is very powerful for tracing out the spatial and velocity structure of the Milky Way's star forming disk.

\section{Methods \& Analysis} \label{sec:meth_anal}

In this section we develop methods and conduct analyses in pursuit of answers to two questions. First, does the previously measured velocity field of the Milky Way in the outer Galaxy allow for illusory spiral structure in the outer Milky Way? Second, if so, is the dense material in the velocity feature actually distributed over a much larger range of distances?

\subsection{Illusory Spiral Structure is Possible} \label{sec:bc_is_possible}

In B71 it was shown that, in the inner Galaxy, significant spiral-like structure could be generated using the velocity field expected from the density-wave theory derived in \cite{1969ApJ...155..721L} with \emph{no density variation at all}. While the idea of a perfectly smooth ISM is inconsistent with essentially every ISM observation, the analysis proved the point that ``the profiles are more sensitive to small variations in the streaming motions than to small variations in the hydrogen density''. This revelation was key to the movement toward studying CO as a tracer of spiral arm structure rather than \hi  (R. Benjamin, priv. comm). 

\begin{figure*}[ht!]
\includegraphics[scale=0.7]{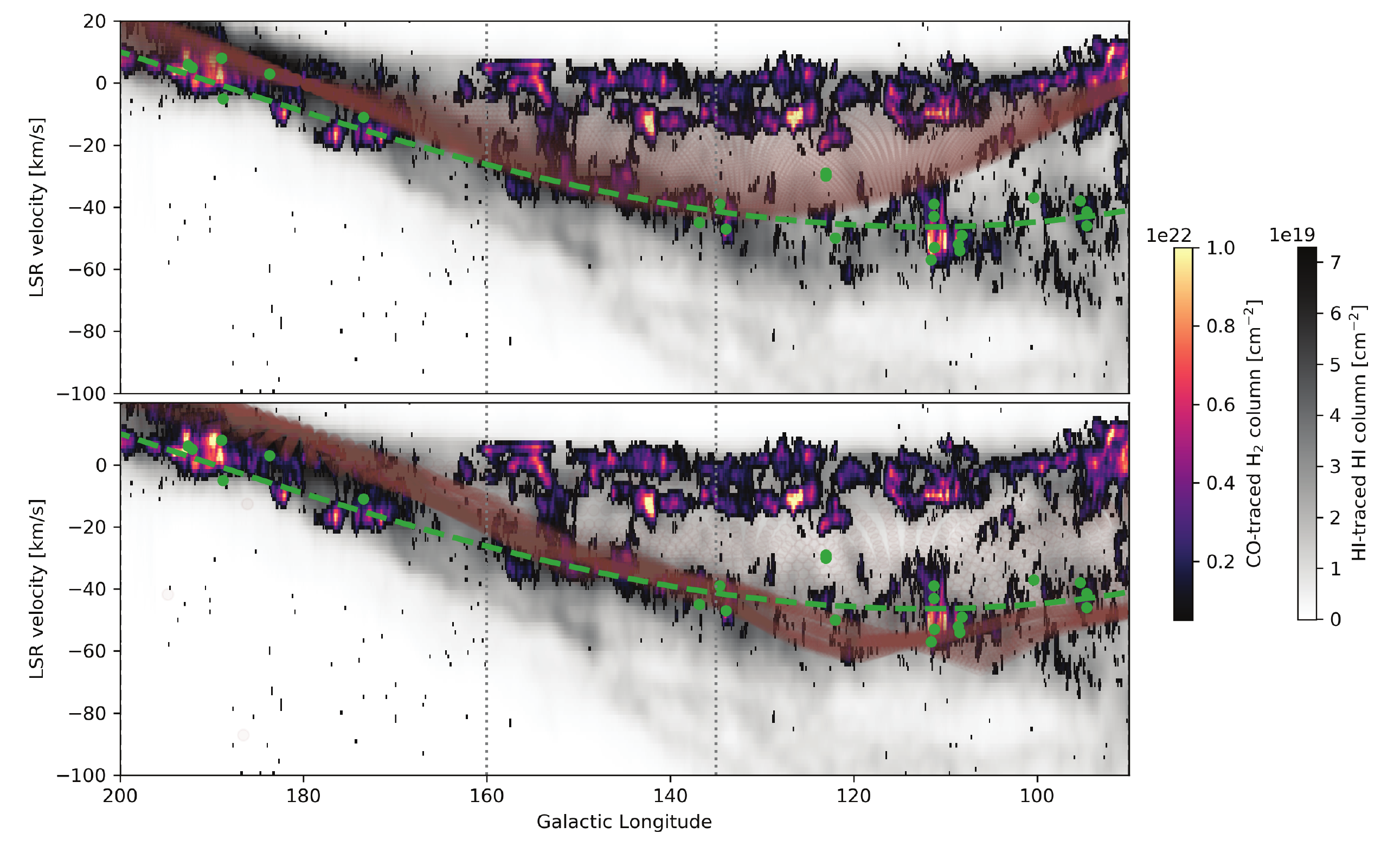}
\caption{The Galactic longitude--velocity diagram in the Galactic midplane ($b=0$). For each panel \hi  column is shown in grayscale and CO column is shown in ``magma'' color map. The bounds of the regions we focus on in \S \ref{sec:bc_is_real} are marked in dotted gray lines. The velocity track of the purported Perseus arm is shown in dashed green from \cite{Choi:2014jg}. Masers from Reid14 and Sakai19 are shown as green dots. Overplotted in brown is a representation of a perfectly smooth ISM. The brown points are smoothly distributed over the Galactic disk within 4 kpc, projected using a flat rotation curve (top) and the velocity field derived in \cite{Tchernyshyov:2018bk} (bottom). This figure shows that track of the Perseus arm can be generated by the velocity field derived in \cite{Tchernyshyov:2018bk} without density fluctuations. \label{fig:bc_is_possible}}
\end{figure*}

To verify this result, and thus demonstrate that velocity crowding in the Milky Way disk is possible, we apply the disk velocity field results of TPZ18 to a perfectly smooth disk. We lay down a smooth grid of points across the second quadrant, out to a distance of 4 kpc where we believe the TPZ18 map to be largely complete and accurate. Then we perform two experiments. First, we use a standard flat, 220 km s$^{-1}$ rotation curve to place each of these points at a velocity, shown in panel 1 of Figure \ref{fig:bc_is_possible}. Then, we use the radial velocity field as measured by TPZ18 to place these same smoothly distributed points on a second longitude--velocity diagram, shown in panel 2 of Figure \ref{fig:bc_is_possible}. 

It is quite clear from these synthetic longitude--velocity diagrams that the Perseus arm in the second quadrant would be easily visible in the longitude--velocity diagram given the TPZ18 map of the velocity field and a perfectly smooth Galactic disk. While this ``velocity crowding'' situation provides the opportunity for illusory features in the longitude--velocity diagram, it does not mean that what we see in the maser-free sections of the Perseus arm is illusory. It is certainly possible that dense material also physically exists at these distances, and that the velocity crowding is only piling on lower density, non-star-forming material at these velocities. Indeed, the HMSFRs from Reid14 and various maps of young stars \citep[e.g.][]{Xu:2018kg} show that at least in some regions there do exist star forming regions at the distance of the Perseus arm.

\subsection{Illusory Spiral Structure Impacts the Perseus Arm Identification} \label{sec:bc_is_real}

In \S \ref{sec:bc_is_possible} we have shown that velocity fields observed in the outer Milky Way disk are sufficiently strong to create spiral arms in the longitude--velocity diagram in the absence of any density enhancements in the disk at all. To determine whether dense gas features are being erroneously connected into longer spiral structures in practice, we seek to measure the true distances to molecular clouds thought to reside in kinematic spiral arms. We focus our attention on the region between $l$=135 and $1$=160 at velocities consistent with the Perseus arm.

We focus on this region because it contains no HMSFRs with parallactic distances and no evidence of young stars clustered at the expected distance Reid14, but has a very clear spiral feature in both \hi  and CO gas and is near enough ($\sim$2 kpc) that the G19 reddening map is generally trustworthy. If this feature exists only in velocity space, and not in distance space, we believe there is currently no evidence of a continuous Perseus spiral arm across the second and into the third quadrant of the outer Galaxy.

\begin{figure*}[ht!]
\plotone{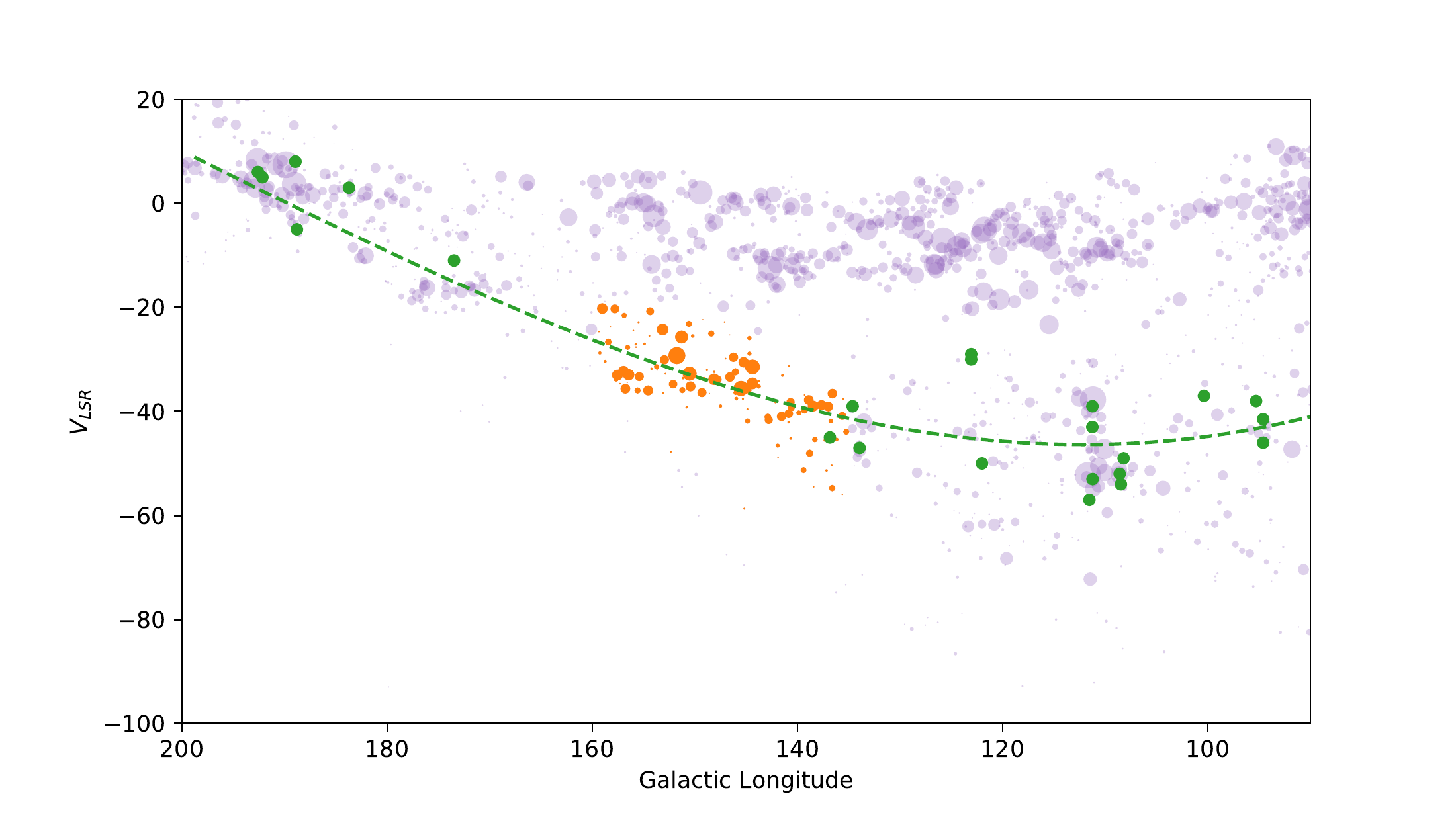}
\caption{Circles represent CO clouds from MML17. CO clouds shown in orange represent the ones we investigate, in purple those not under investigation. Green dots are HMSFRs from Reid14 and Sakai19.\label{fig:mamd_bridge_og}}
\end{figure*}

First we select a sample of clouds from MML17, which are part of the visually identifiable Perseus arm. First we constrain ourselves in longitude: $135^\circ < l < 160^\circ$. Then we make overall constraints on velocity: $-65 {\rm ~km ~s}^{-1} < V_{\rm LSR} < -20 {\rm~ km ~s}^{-1}$. Finally we fine-tune the velocity boundaries to follow the arm in longitude--velocity space:
$V_{\rm LSR}/{\rm ~km ~s}^{-1} < (l - 169^\circ)/^{\circ}$ and $0.6\left(V_{\rm LSR}+35 {\rm~ km ~s}^{-1}\right)/{\rm~ km~ s}^{-1} >  (l - 160^\circ)/^{\circ}$
These constraints result in 114 CO clouds, shown in Figure \ref{fig:mamd_bridge_og}. For each one of these clouds we measure a distance using the G19 map. If the ISM consisted only of these clouds, the procedure for determining their distance would be relatively simple -- measure the average $\Delta {\rm E\left(B-V\right)}\left(d\right)$ along the line of sight and find the peak. Works like \cite{Schlafly:2014hh} and \cite{Zucker:gs} have made very detailed studies of distances to clouds and sub-clouds using some of the techniques in G19, but these have typically been at closer distances or further off the plane. Unfortunately, there are many other structures along the line of sight, including the CO clouds in very strong Local Arm between 0 and -20 km s$^{-1}$ (see figure \ref{fig:bc_is_possible}) and the more diffuse phases of the ISM along the line of sight not detected in DMT01. 

\begin{figure*}[ht!]
\gridline{\fig{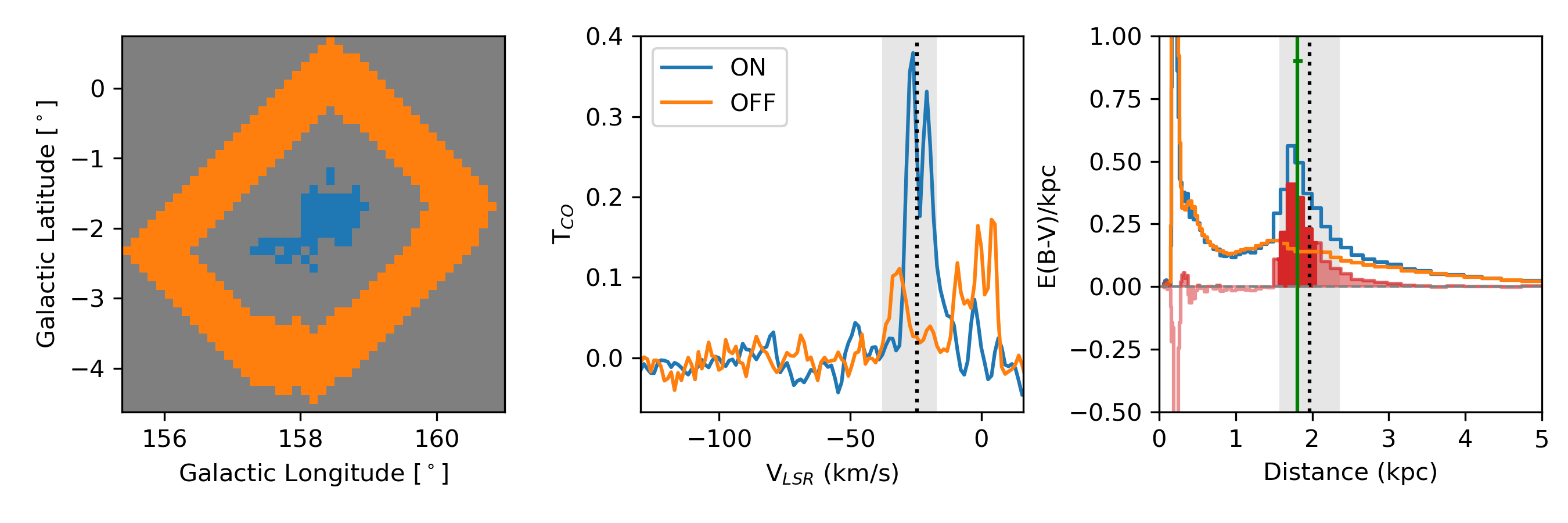}{1.0\textwidth}{}}
\gridline{\fig{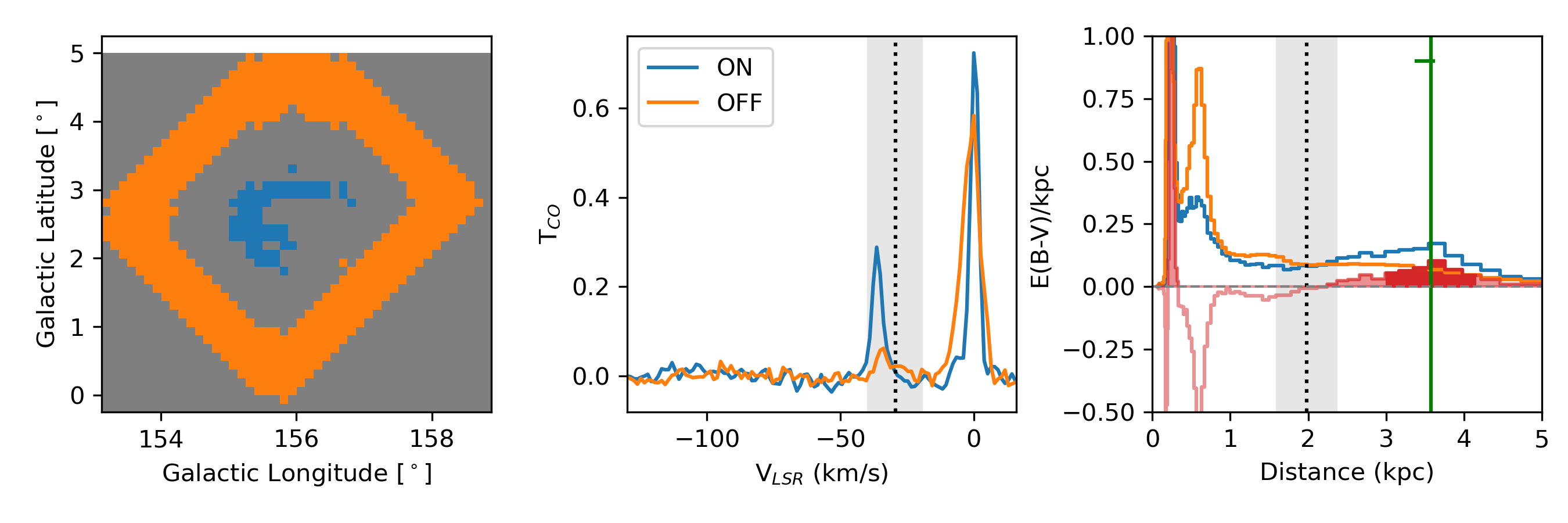}{1.0\textwidth}{}}
\gridline{\fig{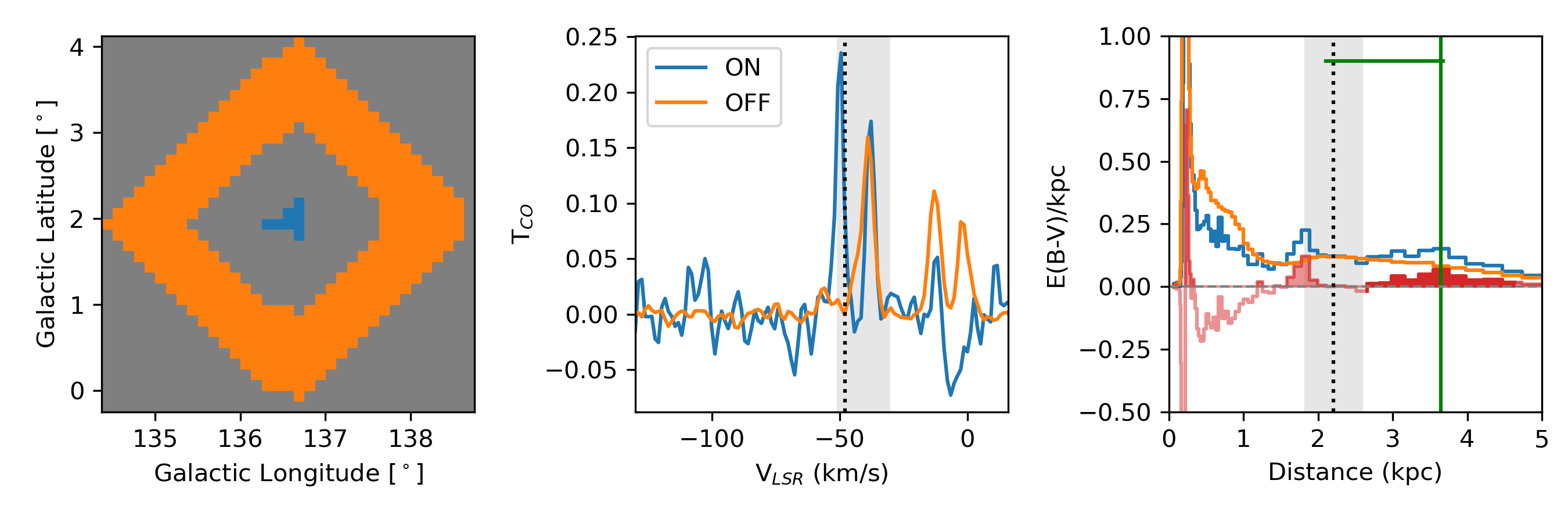}{1.0\textwidth}{}}

\caption{Three clouds demonstrating three main results from the analysis. The left column of panels shows the on region representing the cloud from MML17 (blue) and the ``off'' region in orange. The CO spectra representing these regions are shown in the second column, with the cloud velocity in black dotted line and gray region representing Perseus velocities. The third column shows the average reddening at each distance for the regions in blue and orange, and their difference in red, with the expected distance to the Perseus arm at the dashed black line. The top cloud is an example where the method finds a result that is consistent with expected distance of the arm, the middle cloud is a successful distance at an inconsistent distance, and the bottom cloud is an example of where the method fails.  \label{fig:method_panels}}
\end{figure*}

To mitigate the effect of foreground ISM structures along the line of sight we also choose a control region, which we call an ``off'', consisting of the area of sky more than 7 pixels away from any region of the cloud and less than 16 pixels in the DMT01 CO map (see left panels of Figure \ref{fig:method_panels}). Each pixel in the DMT01 map is 7.5$^\prime$, such that the off region consists of the area roughly 1$^\circ$ to 2$^\circ$ from the cloud in question. For each of the cloud and off region we find all the corresponding dust pixels in G19 that have complete data out to at least 4 kpc. 

For both the cloud and the off region we make a weighting map consisting of total column of material along the line of sight excepting the cloud itself. To do this we use the fiducial column density conversion factors for 21-cm \hi  and microwave CO 1-0 transitions \cite{2013ARA&A..51..207B}

:
\begin{equation}
N\left(\rm H~ \sc{I} \right) = 1.8 \times 10^{18} \frac{{\rm T_B} d\nu}{\rm K~ km~ s^{-1}}
\end{equation}
and
\begin{equation}
X\left(\rm CO\right) = 2.0 \times 10^{20} \frac{{\rm T_B} d\nu}{\rm K~ km~ s^{-1}}
\end{equation}
apply them to the CO and \hi  maps, sum them, and subtract the contribution from the cloud according to MML17. These maps allow us to weight our average measurement of the dust for the clouds and the offs by the inverse of the unrelated ISM column. This weighting allows us to focus on sightlines that are less contaminated by intervening and unrelated material, though in practice we find using an unweighted average does not change the conclusions of this work.

We look to estimate the distance to each cloud using the dust reddening measurements of G19. To compute this distance we construct the difference between our on-cloud weighted $\Delta {\rm E\left(B-V\right)}\left(d\right)$ and the off-cloud weighted $\Delta {\rm E\left(B-V\right)}\left(d\right)$. In the very near field, $d < 0.7$ kpc, we find that it is impossible to fully subtract off the effects of the Local Arm material, which is very strong and quite variable. Thus, we focus our attention on the range 0.7 kpc $< d <$ 5 kpc in our analysis. For each cloud we find the interval over this distance range that captures 68.3\% of the nonnegative cloud-off $\Delta {\rm E\left(B-V\right)}\left(d\right)$ that covers the least distance, and use the center of this range as our distance estimator (see right panels of Figure \ref{fig:method_panels}). We find this method is much more resilient to the complex, non-Gaussian distributions that are the result of the complex intervening ISM than e.g. first moment measurements. 

To measure errors we turn to block bootstrapping. Standard bootstrap analysis \citep{efron1982jackknife} is extremely useful for situations where errors on samples are uncorrelated but otherwise quite complex. Unfortunately, while the errors in distance for this measurement are indeed complex (being totally dominated by intervening clouds rather than measurement noise) they are by no means independent: two adjacent pixels are very likely to be contaminated by the same intervening cloud. Thus, standard bootstrapping is both technically inapplicable and practically inaccurate. For this reason we turn to block bootstrapping, a procedure of resampling with replacement, but for subregions (blocks), rather than individual pixels \cite{Hall:1985ie}.

To make subregions for a given region (cloud or off) we first find the average position of the region in Galactic coordinates. Then we compute a clock-angle on the surface of the sky for each of the pixels in the region around this center position. Finally, we divided the region up into blocks based on the pixels clock-angle rank, with the largest block having no more than one extra pixel than the smallest block. This creates largely contiguous blocks of pixels of equal size. We use 12 blocks for each region. As we perform this block bootstrap we also bootstrap on the five realizations of the dust map. We find that the map realization bootstrapping adds very little to the overall errors, consistent with our expectation that intervening clouds, and not measurement noise, are our largest source of uncertainty.

\begin{figure*}[ht!]
\includegraphics[scale=0.75]{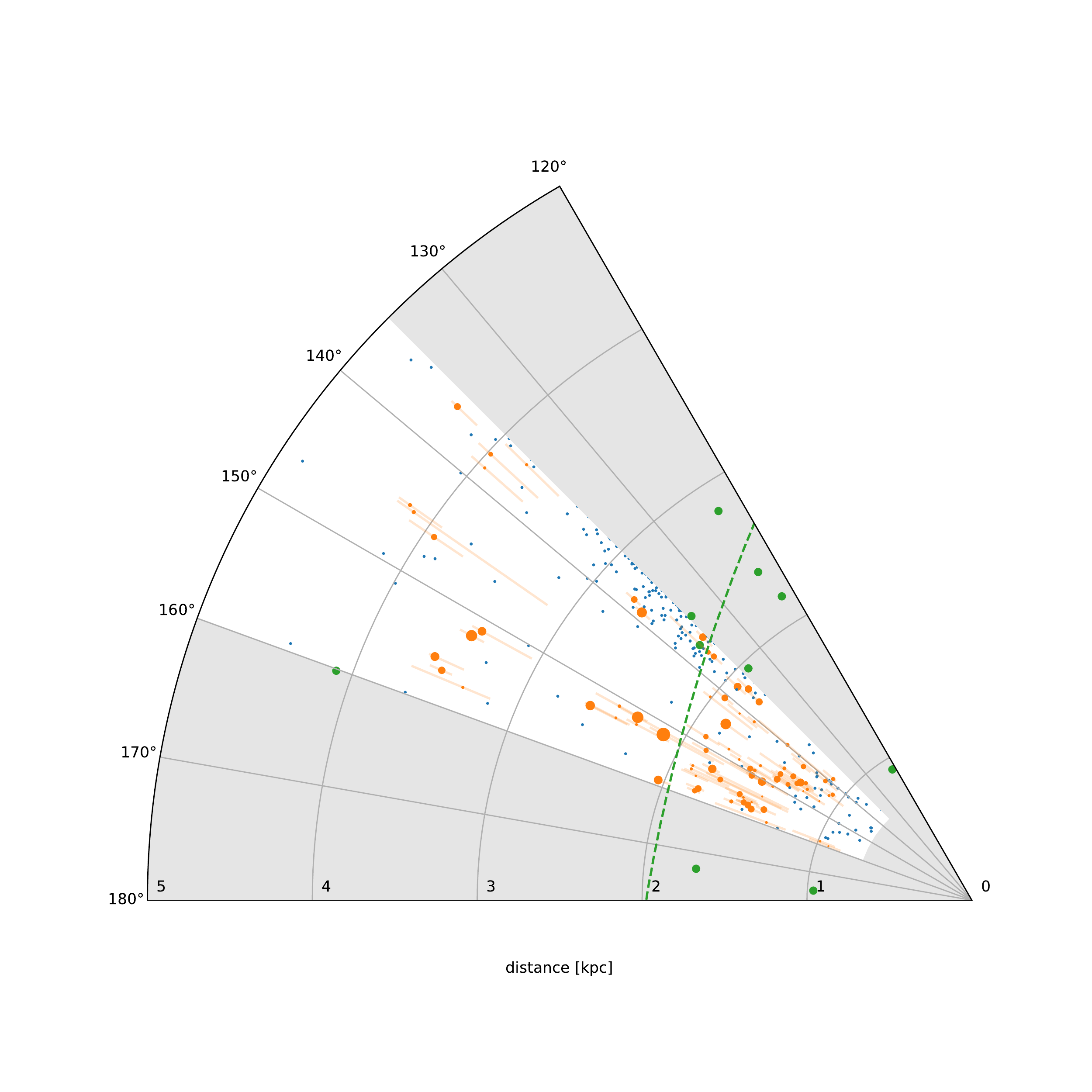}
\caption{Distribution of 81 MML17 clouds determined by reddening analysis, shown in orange. Gray regions are not examined by the method. Dashed green line represents the expected position of the Perseus arm from \cite{Choi:2014jg}. HMSFRs from Reid14 and Sakai19 are shown as green dots. In blue we show the stars from \cite{Xu:2018kg} that have parallaxes detected at $>$5 $\sigma$ within this region. A number of features match between the stars and clouds. The clouds, like the stars, do not seem to bear any relation to the arm model, except very near the HMSFRs at $l\sim140^\circ$ as expected. \label{fig:wedge}}
\end{figure*}

\begin{figure*}[ht!]
\includegraphics[scale=0.75]{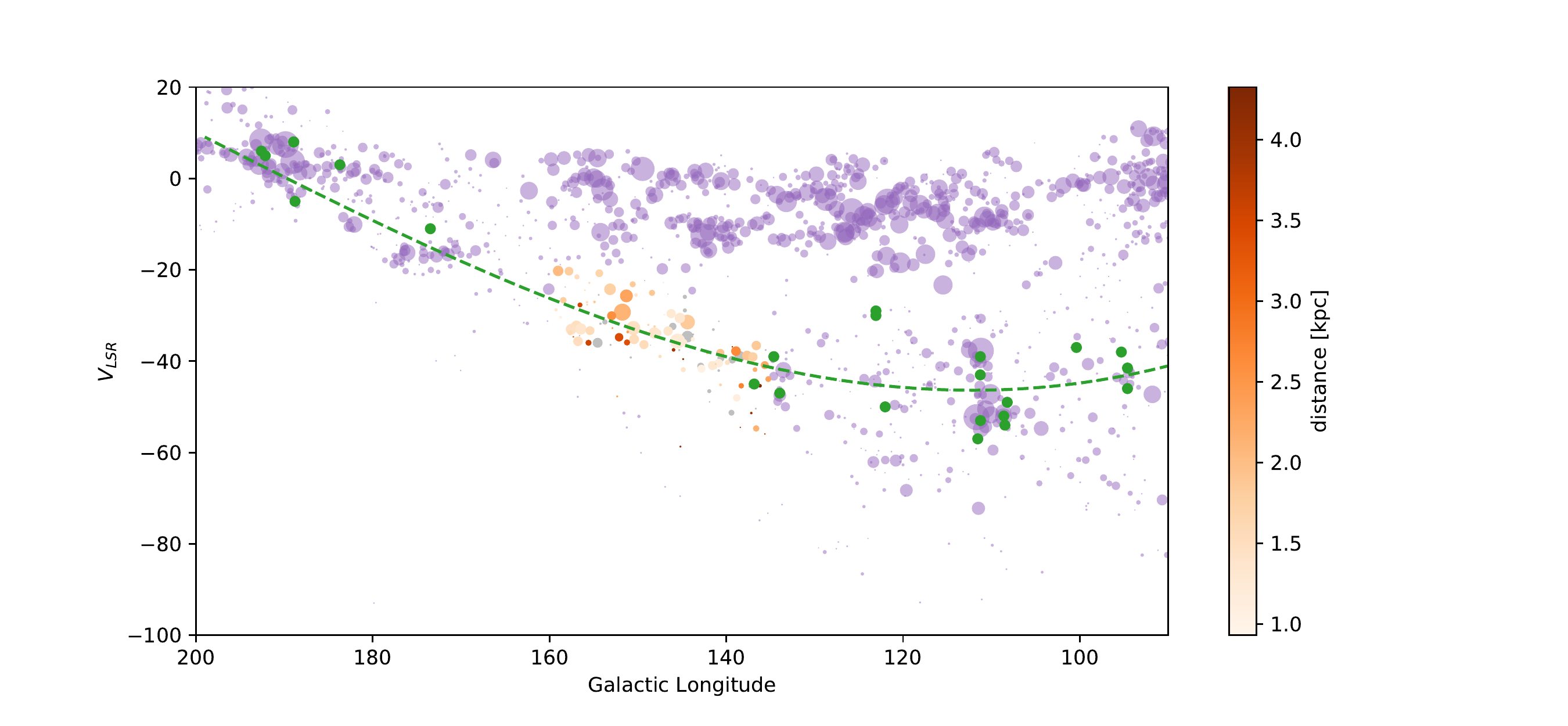}
\caption{Circles are CO clouds from MML17. Color bar represents the distance found through reddening the analysis. Green dots are Reid14 and Sakai19 HMSFRs. The green dashed line is the Perseus arm model from \cite{Choi:2014jg}. Clouds that are very near each other in the longitude--velocity diagram often have very different distances \label{fig:mamd_bridge}}
\end{figure*}

While we have built a system to find cloud distances that is robust as possible in the presence of intervening material, some clouds are too small or too obscured to find with this method. 
First, we reject 26 additional clouds for having a $\pm1 \sigma$ error range that is more than a factor of 1.5 in distance. Second, we reject 8 clouds for having very heavy-tailed dust profiles, defined by having a 68.3\% ($\pm 1\sigma$) interval 3.5 times larger than their 38.3\% ($\pm 0.5\sigma$) interval or having a 86.6\% ($\pm 1.5\sigma$) interval 3.5 times larger than their 68.3\% interval. This leaves us a sample of 81 of 114 initial clouds which we report in Table \ref{tab:clouds} and map in Figure \ref{fig:wedge}.  We note that this procedure makes it clear that we are biased away from finding the most distant clouds, as they will tend to be fainter and smaller overall. We expect that this sample is thus more complete closer to the Sun. We see in Figure \ref{fig:wedge} that the bulk of CO clouds at the Perseus arm velocity in this longitude range are not consistent with being at the distance of the Perseus arm. The longitude--velocity diagram of the Perseus arm is shown with distances assigned in Figure \ref{fig:mamd_bridge}, clearly demonstrating a very wide range of distances for clouds very near each other in velocity space.

\startlongtable
\begin{deluxetable*}{ccccccccc}
\tablecaption{Clouds with measured dust distances. \label{tab:clouds}}
\tablecolumns{5}
\tablehead{
\colhead{Cloud} & \colhead{MML17 Cloud} & \colhead{$l$} & \colhead{$b$} & \colhead{V$_{\rm LSR}$}& \colhead{Area} & \colhead{Av. Flux} &\colhead{Dust Distance}\\
\colhead{Number} & \colhead{Number} & \colhead{[$^\circ$]} & \colhead{[$^\circ$]}& \colhead{[km s$^{-1}$]} & \colhead{[deg$^2$]} & \colhead{[K km s$^{-1}$]}  & \colhead{[kpc]}
}
\startdata
1 & 2168 & 137.65 & 1.41 & -38.8 & 1.31 & 2.8 & $1.92^{+0.09}_{-0.06}$\\
2 & 2178 & 136.60 & 1.17 & -36.6 & 0.78 & 4.4 & $1.86^{+0.09}_{-0.03}$\\
3 & 2182 & 148.13 & 0.25 & -33.8 & 1.92 & 2.6 & $1.39^{+0.03}_{-0.08}$\\
4 & 2190 & 144.39 & -1.20 & -31.4 & 2.27 & 3.7 & $1.84^{+0.02}_{-0.16}$\\
5 & 2201 & 151.76 & -1.19 & -29.3 & 2.62 & 4.2 & $2.12^{+0.09}_{-0.34}$\\
6 & 2702 & 136.75 & 1.10 & -41.8 & 0.23 & 3.8 & $2.20^{+0.31}_{-0.06}$\\
7 & 2712 & 137.00 & 1.37 & -39.1 & 1.03 & 3.4 & $1.76^{+0.09}_{-0.01}$\\
8 & 2731 & 145.50 & 0.86 & -35.6 & 2.06 & 4.2 & $1.26^{+0.03}_{-0.09}$\\
9 & 2736 & 150.52 & -0.94 & -32.7 & 3.20 & 2.4 & $1.46^{+0.08}_{-0.06}$\\
10 & 2749 & 151.30 & -0.53 & -25.7 & 1.16 & 5.5 & $2.31^{+0.12}_{-0.59}$\\
11 & 2757 & 150.58 & -0.88 & -23.2 & 0.41 & 3.6 & $1.85^{+0.09}_{-0.35}$\\
12 & 3397 & 140.85 & -1.17 & -40.4 & 0.66 & 4.4 & $1.15^{+0.26}_{-0.09}$\\
13 & 3398 & 141.56 & -1.22 & -41.0 & 1.86 & 1.7 & $1.30^{+0.08}_{-0.05}$\\
14 & 3417 & 140.68 & 0.81 & -38.2 & 1.30 & 2.0 & $1.94^{+0.09}_{-0.06}$\\
15 & 3422 & 156.78 & -1.85 & -35.6 & 1.44 & 2.5 & $1.51^{+0.03}_{-0.00}$\\
16 & 3429 & 144.80 & 1.01 & -35.9 & 0.41 & 4.6 & $1.23^{+0.08}_{-0.04}$\\
17 & 3430 & 146.03 & 1.01 & -36.4 & 0.17 & 4.6 & $1.20^{+0.16}_{-0.07}$\\
18 & 3436 & 156.95 & -2.94 & -32.3 & 1.52 & 2.9 & $1.48^{+0.07}_{-0.04}$\\
19 & 3440 & 157.55 & 0.85 & -33.0 & 1.91 & 2.4 & $1.45^{+0.03}_{-0.06}$\\
20 & 3444 & 145.24 & -1.08 & -30.5 & 1.69 & 2.5 & $1.32^{+0.24}_{-0.14}$\\
21 & 3454 & 146.24 & 1.25 & -29.6 & 1.09 & 3.0 & $1.28^{+0.03}_{-0.09}$\\
22 & 3477 & 154.37 & -1.15 & -20.7 & 0.91 & 2.6 & $1.69^{+0.09}_{-0.45}$\\
23 & 4324 & 135.64 & -0.13 & -40.9 & 1.66 & 1.5 & $2.28^{+0.05}_{-0.07}$\\
24 & 4338 & 142.80 & -0.96 & -41.7 & 1.80 & 1.4 & $1.06^{+0.05}_{-0.02}$\\
25 & 4356 & 144.87 & 1.56 & -41.9 & 0.55 & 1.9 & $1.39^{+0.03}_{-0.06}$\\
26 & 4360 & 138.90 & -0.41 & -37.8 & 2.92 & 1.2 & $2.66^{+0.04}_{-0.06}$\\
27 & 4363 & 139.87 & 0.29 & -40.3 & 0.67 & 1.5 & $1.46^{+0.22}_{-0.31}$\\
28 & 4386 & 150.43 & 0.42 & -35.2 & 1.81 & 2.1 & $1.53^{+0.03}_{-0.04}$\\
29 & 4388 & 151.23 & 0.73 & -35.9 & 1.08 & 1.4 & $3.39^{+0.06}_{-0.34}$\\
30 & 4394 & 152.13 & 1.57 & -34.8 & 2.05 & 1.3 & $3.43^{+0.07}_{-0.08}$\\
31 & 4396 & 155.59 & 2.54 & -36.0 & 0.97 & 1.5 & $3.58^{+0.03}_{-0.19}$\\
32 & 4402 & 147.78 & -0.53 & -33.9 & 1.25 & 1.9 & $1.32^{+0.10}_{-0.07}$\\
33 & 4403 & 156.45 & -0.37 & -33.0 & 3.52 & 1.4 & $1.38^{+0.05}_{-0.07}$\\
34 & 4404 & 151.16 & -0.32 & -33.6 & 0.11 & 2.4 & $2.44^{+0.16}_{-0.19}$\\
35 & 4407 & 149.11 & -0.16 & -33.8 & 0.41 & 1.8 & $1.53^{+0.03}_{-0.15}$\\
36 & 4409 & 149.31 & 0.75 & -36.4 & 1.98 & 1.6 & $1.56^{+0.05}_{-0.04}$\\
37 & 4410 & 146.58 & 0.47 & -33.4 & 1.61 & 2.1 & $1.39^{+0.03}_{-0.12}$\\
38 & 4414 & 155.42 & 1.35 & -33.3 & 2.11 & 1.4 & $1.55^{+0.36}_{-0.10}$\\
39 & 4417 & 157.68 & 1.80 & -33.8 & 0.78 & 1.2 & $1.58^{+0.04}_{-0.06}$\\
40 & 4427 & 152.97 & -0.85 & -30.1 & 1.66 & 1.9 & $2.60^{+0.03}_{-0.24}$\\
41 & 4461 & 158.44 & -1.98 & -26.7 & 0.91 & 1.9 & $1.81^{+0.04}_{-0.03}$\\
42 & 4463 & 148.40 & -1.58 & -25.0 & 1.05 & 1.5 & $1.89^{+0.13}_{-0.09}$\\
43 & 4476 & 153.15 & -1.55 & -24.3 & 2.08 & 2.5 & $1.76^{+0.06}_{-0.04}$\\
44 & 4494 & 157.80 & -2.24 & -20.3 & 1.44 & 2.1 & $1.79^{+0.07}_{-0.03}$\\
45 & 4495 & 159.02 & -2.30 & -20.2 & 1.88 & 2.3 & $2.04^{+0.01}_{-0.08}$\\
46 & 4496 & 156.89 & -2.35 & -21.5 & 0.55 & 2.0 & $1.51^{+0.04}_{-0.03}$\\
47 & 5877 & 136.62 & 0.29 & -54.7 & 1.09 & 1.5 & $2.15^{+0.07}_{-0.06}$\\
48 & 5892 & 145.19 & 2.83 & -58.7 & 0.14 & 1.0 & $4.12^{+0.12}_{-0.98}$\\
49 & 5909 & 135.63 & 1.81 & -55.9 & 0.08 & 0.9 & $3.78^{+0.17}_{-0.27}$\\
50 & 5935 & 138.82 & -1.83 & -48.0 & 1.19 & 1.8 & $1.12^{+0.08}_{-0.01}$\\
51 & 5954 & 137.17 & 3.12 & -51.4 & 0.20 & 1.2 & $3.98^{+0.09}_{-0.38}$\\
52 & 5960 & 152.34 & -0.70 & -47.7 & 0.16 & 0.9 & $2.30^{+0.06}_{-0.22}$\\
53 & 5981 & 138.30 & -0.75 & -45.4 & 1.02 & 1.2 & $2.74^{+0.06}_{-0.04}$\\
54 & 5988 & 140.66 & 0.32 & -45.2 & 0.28 & 1.1 & $1.71^{+0.04}_{-0.10}$\\
55 & 6000 & 136.18 & 2.12 & -45.4 & 0.48 & 1.1 & $4.32^{+0.04}_{-0.16}$\\
56 & 6047 & 141.21 & 1.58 & -40.6 & 0.09 & 1.0 & $1.81^{+0.08}_{-0.13}$\\
57 & 6057 & 147.50 & -2.28 & -39.0 & 0.42 & 1.0 & $1.52^{+0.09}_{-0.27}$\\
58 & 6070 & 145.96 & 2.62 & -37.5 & 0.47 & 1.1 & $3.94^{+0.18}_{-0.20}$\\
59 & 6074 & 142.15 & 3.31 & -38.2 & 0.27 & 0.8 & $2.01^{+0.05}_{-0.32}$\\
60 & 6078 & 143.78 & -3.38 & -35.2 & 0.17 & 4.2 & $1.07^{+0.07}_{-0.10}$\\
61 & 6097 & 144.88 & 2.34 & -39.5 & 0.12 & 1.1 & $4.17^{+0.08}_{-0.23}$\\
62 & 6103 & 148.12 & -2.35 & -32.4 & 0.12 & 1.7 & $1.74^{+0.07}_{-0.08}$\\
63 & 6125 & 148.82 & 2.11 & -32.0 & 0.12 & 1.5 & $1.65^{+0.06}_{-0.22}$\\
64 & 6131 & 157.30 & 2.88 & -34.7 & 0.08 & 1.2 & $3.35^{+0.33}_{-0.17}$\\
65 & 6135 & 154.24 & -0.91 & -31.8 & 0.23 & 1.0 & $1.88^{+0.01}_{-0.59}$\\
66 & 6142 & 158.75 & 0.01 & -30.4 & 0.39 & 0.9 & $0.99^{+0.17}_{-0.09}$\\
67 & 6143 & 152.87 & 0.22 & -32.8 & 0.14 & 0.8 & $2.43^{+0.16}_{-0.09}$\\
68 & 6155 & 156.55 & 2.45 & -27.7 & 0.94 & 1.0 & $3.50^{+0.07}_{-0.06}$\\
69 & 6163 & 155.74 & -2.73 & -27.6 & 0.08 & 1.2 & $1.84^{+0.09}_{-0.08}$\\
70 & 6164 & 147.01 & -2.44 & -29.8 & 0.09 & 1.2 & $1.10^{+0.10}_{-0.03}$\\
71 & 6165 & 154.91 & -2.03 & -27.0 & 0.22 & 1.2 & $1.88^{+-0.01}_{-0.25}$\\
72 & 6170 & 155.76 & -1.22 & -27.1 & 0.30 & 0.8 & $1.52^{+0.03}_{-0.09}$\\
73 & 6171 & 159.26 & -0.85 & -28.8 & 0.53 & 0.8 & $1.33^{+0.33}_{-0.12}$\\
74 & 6183 & 150.21 & -2.51 & -25.5 & 0.38 & 1.3 & $1.39^{+0.29}_{-0.10}$\\
75 & 6185 & 156.00 & -2.41 & -24.5 & 0.09 & 1.1 & $1.46^{+0.14}_{-0.04}$\\
76 & 6222 & 159.36 & 0.88 & -24.7 & 0.09 & 0.9 & $0.93^{+0.12}_{-0.07}$\\
77 & 6238 & 147.11 & -2.34 & -22.8 & 0.08 & 1.1 & $1.22^{+0.23}_{-0.10}$\\
78 & 6240 & 155.49 & -1.88 & -22.9 & 0.19 & 0.9 & $1.55^{+0.09}_{-0.01}$\\
79 & 9099 & 138.41 & 3.42 & -54.5 & 0.08 & 0.8 & $3.95^{+0.10}_{-0.30}$\\
80 & 9139 & 153.68 & 1.54 & -32.0 & 0.08 & 0.8 & $1.42^{+0.32}_{-0.17}$\\
\enddata
\end{deluxetable*}

\section{Discussion} \label{sec:disc}

\begin{figure*}[ht!]
\includegraphics[scale=0.74]{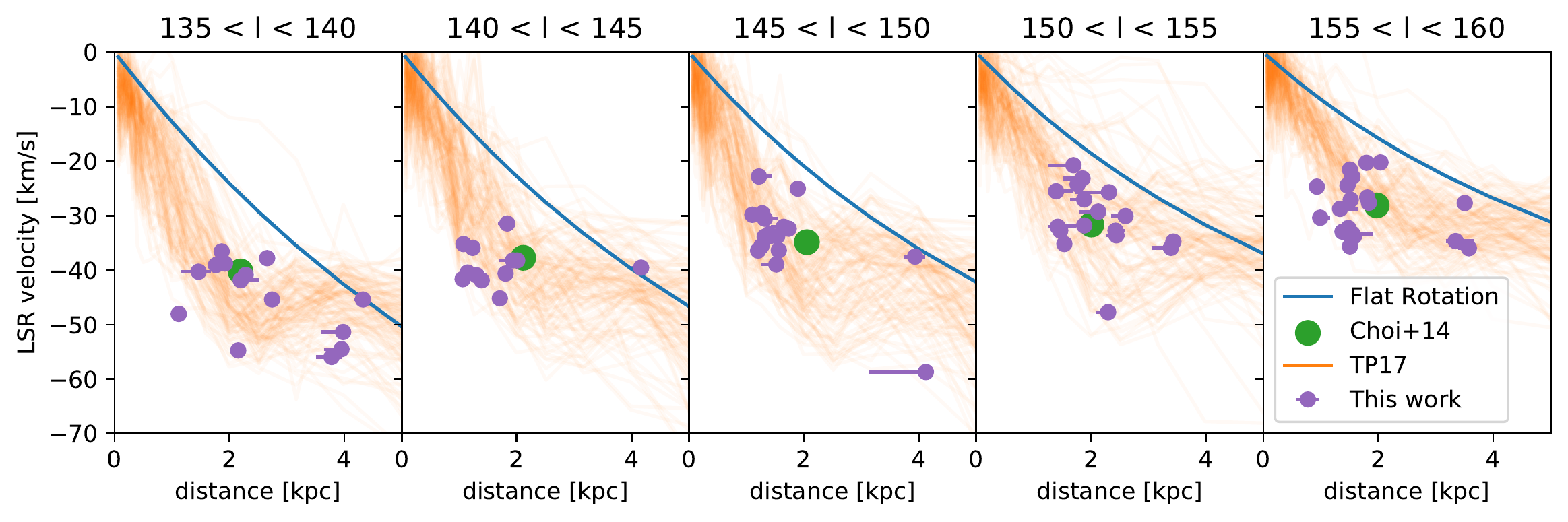}
\caption{Distance-velocity diagrams for five degree sections of the Galactic plane. The standard 220 km s$^{-1}$ flat rotation curve is shown in blue. The model of the Perseus arm presented in \cite{Choi:2014jg} is shown as a green dot for the central longitude of the region. The orange lines represent the results of TP17. These distance--velocity lines are a grid of samples across the longitude selection and within -2.5$^{\circ} < b < 2.5^{\circ}$, spaced at 0.5$^{\circ}$ intervals. The purple dots and error bars are the results of our analysis presented in Table \ref{tab:clouds}.
\label{fig:DV_for_BC}}
\end{figure*}

Section \ref{sec:bc_is_possible} and Figure \ref{fig:bc_is_possible} show that the velocity field toward the Perseus Arm in the second quadrant of the Milky Way is able to produce the structure-emulating velocity crowding effect discussed by \cite{1971A&A....10...76B}. From \S \ref{sec:bc_is_real} and Figure \ref{fig:wedge} we can see that this velocity crowding is indeed solely responsibly for the bridge of molecular clouds that connect star forming regions at $l\simeq140^\circ$ to those in the anti-center in the longitude--velocity diagram. From this analysis we find no evidence for any kind of overdensity of molecular clouds at the distance of the Perseus arm over this range in longitude. The analyses conducted in Section \ref{sec:bc_is_possible} and Section \ref{sec:bc_is_real} produce consistent results, in that clouds in the Perseus component of the l-v diagram are actually distributed over a very large range of distances. Indeed, the clouds generally line up with the sample of O and B stars described in \cite{Xu:2018kg}. This result is consistent with the lack of evidence for a stellar overdensity in this region in \citet{Poggio:wd}. Further, the wedge of molecular clouds we find closer than the purported Perseus arm is somewhat consistent with the ``Cepheus Spur'' of young stars highlighted most recently in \citet{2021MNRAS.504.2968P}, although we note the overlap is not complete. In Figure \ref{fig:mamd_bridge} the impact of velocity crowding is made clear; clouds with very different distances can be put at very similar places in the longitude--velocity diagram. The velocity bridge we have interrogated here is, to our knowledge, the only connection between the star forming region that ends at $l\simeq135^\circ$ and the cluster near Galactic anti-center, $l\simeq180^\circ$. Additionally, we show the cloud distance results in distance--velocity space broken up over five regions of Galactic longitude in Figure \ref{fig:DV_for_BC}. In this figure we include the velocities derived in TP17 to illustrate both how the final results  of this work consistent with TP17 and how many distances correspond to the same narrow range of velocities over many sightlines. This figure very clearly shows that it is very difficult to directly relate a velocity to a distance over this area of sky in the velocity range we are examining. At velocities above about -30 km s$^{-1}$ the TP17 curves show a distance consistently closer than the purported track of the Perseus arm, with velocities closer to zero typically closer to the Sun, but at lower velocities there is essentially no consistently measurable relationship between distance and velocity.

\begin{figure*}[ht!]
\plotone{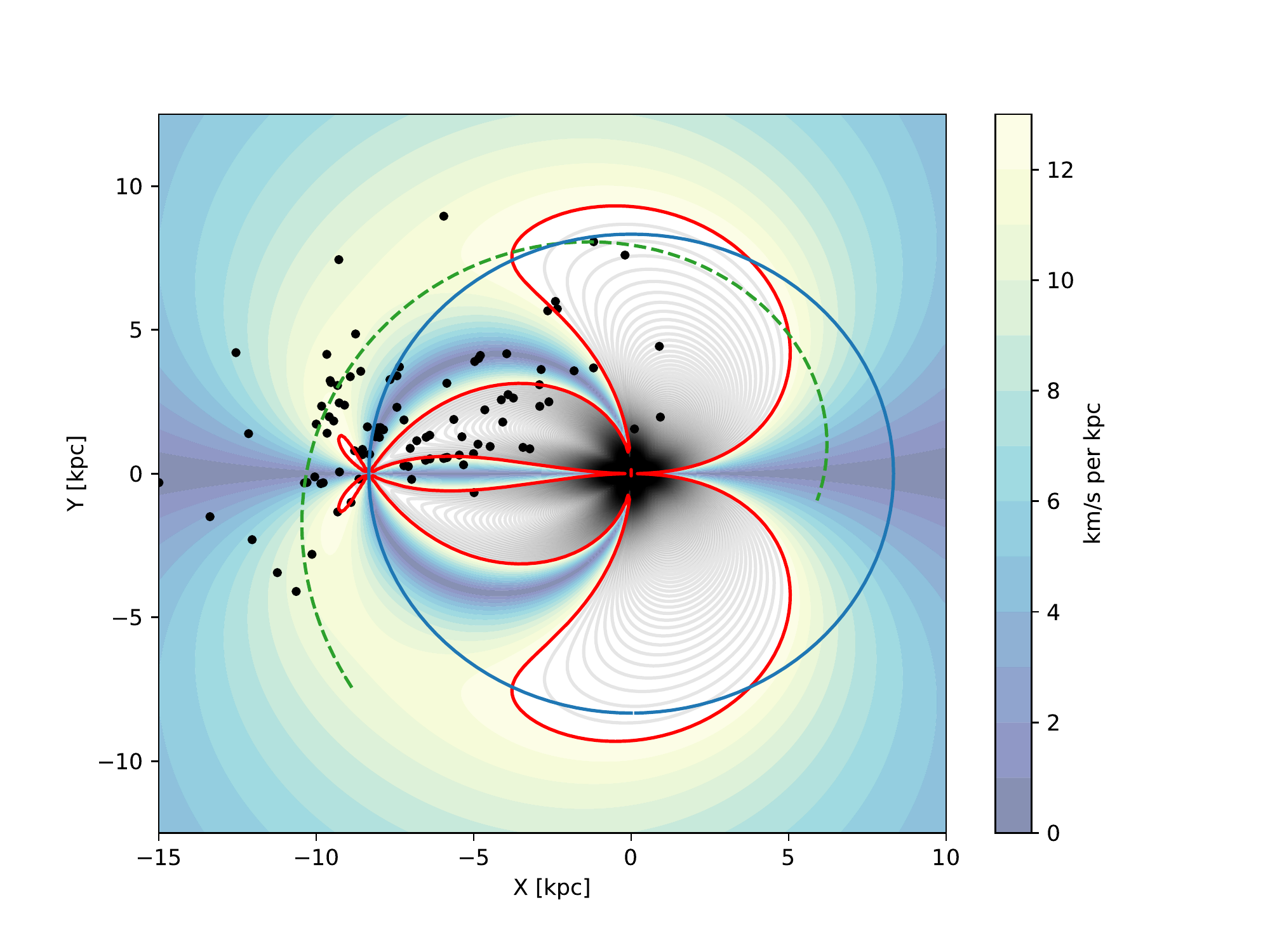}
\caption{A top-view of the Galactic disk, with the sun to the left and Galactic center at (0, 0). The arc of the Perseus arm  defined in \cite{Choi:2014jg} is shown in green. Regions with a shallower distance--velocity slope expected from a flat rotation curve than seen toward Perseus shown in color scale, regions with steeper slope shown in gray scale, separated by red line. Masers from Reid14 are shown in black. Almost all of the outer disk could be impacted at a level similar to the region we consider, as could inner-disk regions near the tangency points and on the far side of the Galaxy. This diagram implies that most connections we draw between masers using l-v diagram to form spiral arms are subject to ``Burton's Curse''.  \label{fig:area_impact}}
\end{figure*}

To understand the scope of problem this effect presents to our disk-mapping efforts across the whole of the disk we chart the steepness of the distance--velocity relationship for an assumed flat, 220 km s$^{-1}$ disk in Figure \ref{fig:area_impact}. The red line shows the boundary of 13 km s$^{-1}$ / kpc, about the steepest expected gradient in the region where we see a velocity crowding effect generate the illusion in the longitude--velocity diagram. This map shows that nearly the entire outer Galaxy and about half of the inner Galaxy are susceptible to these illusory features given bulk flows of the scale seen toward Perseus. This provides a serious warning to investigators using the longitude--velocity diagram to link star-forming features in large swathes of the Milky Way disk.

\section{Conclusion} \label{sec:conc}

In this work we have shown that, as originally proposed in B71, real velocity deviations from a flat Galactic rotation curve can a have significant impact on our ability to discern spiral structure in the Galactic longitude--velocity diagram (Figure \ref{fig:bc_is_possible}). In particular, we demonstrated that the longitude--velocity bridge of molecular gas connecting the star forming regions at $l = 135^\circ$ and $l = 180^\circ$ is entirely an artifact of these velocity deviations (Figure \ref{fig:wedge}). The amplitude of these velocity variations can create confusion in the longitude velocity diagram for the bulk of the Galactic disk (Figure \ref{fig:area_impact}).

We believe these results have significant implications for the study of Galactic structure. On the smaller scale, the Perseus arm star is often used as a proving ground for theories of star formation. This work shows that one cannot simply lump gas together at Perseus arm velocities in the second quadrant and expect to be probing only a spiral arm; gas at the Perseus arm velocity can extend over at least 3 kpc in radial distance. On larger scales, we believe this work implies that Galactic spiral arm identification, especially when many discrete star-forming regions are connected using the longitude--velocity diagram, requires true 3D evidence. Luckily, recent dust tomography work (e.g. G19) has been able to probe gas out to many kpc from the Sun. To extend this work, and reveal the true spiral structure of the Galaxy, we will need even deeper observations and more sophisticated stellar and dust inferences schema. We hope this motivates the pursuit of deep and precise observations by upcoming large-angle imaging surveys, such as those conducted with the Vera Rubin Observatory and the Nancy Grace Roman Space Telescope.
\\
\begin{center}
ACKNOWLEDGEMENTS
\end{center}

This work was supported in part by the US National Science Foundation under grant 1616177. The authors profusely thank Bob Benjamin for long discussions, guidance, and wisdom without which this work would not have been possible. The authors acknowledge Paris-Saclay University’s Institut Pascal programs ``The Self-Organized Star Formation Process'' and ``The Grand Cascade'' and the Interstellar Institute for hosting discussions that nourished the development of the ideas behind this work.  The authors thank an anonymous referee whose report has significantly enhanced this work.  JEGP thanks Eddie Schlafly for pointing out the Perseus arm discrepancy in the dust maps to begin with.

\begin{appendix}

\section{Corrected and augmented version of the MML17 molecular clouds catalog}

The analysis in this paper relies on a corrected version of the molecular cloud catalog of \cite{MivilleDeschenes:2017go}. 

\subsection{Astrometry error correction} 

An error in the coordinates of the cloud catalog published in MML17 has been identified after publication: GLAT is offset by one beam (7.5 arcmin), and the central velocity is offset by one channel (1.3 km/s). 
Correcting this astrometry error resulted in a slight increase in the number of clouds for which a kinematic distance could be estimated, from 8107 to 8246 clouds. The current study uses this corrected version of the catalog. 

\subsection{Extended products}

The original catalog provided in MML17 contains only clouds for which a kinematic distance could be estimated. These clouds correspond to 89\% of the total CO emission, while the whole cloud data set contains 98\% of the emission. To extend the original product we are providing a catalog that contains the full set of 9710 clouds that encapsulate 98\% of the $^{12}$CO (J=1-0) emission. Clouds for which a kinematic distance could not be estimated do not have mass, density, or physical size estimates. 

In addition to this extended cloud catalog, we used the full set of Gaussian components used to describe the $^{12}$CO emission of the Milky Way disk to compute a series of extended products. These include
\begin{itemize}
    \item The whole Gaussian data set used to described the CO emission. 
    \item All 9710 reconstructed CO emission position-position-velocity cubes, one per cloud
    \item 2D maps of the number of clouds and the number of Gaussian on each line of sight.
    \item A cube providing the list of Cloud ID at each sky position
    \item A reconstruction of the whole $^{12}$CO data cube, only based on the Gaussian components.
\end{itemize}

All these products are publicly available on Dataverse at \dataset[10.7910/DVN/QR9CFW]{https://doi.org/10.7910/DVN/QR9CFW}.

\end{appendix}

%\bibliography{all.bib,extra.bib}

\end{document}